# Can absolute freedom save quantum mechanics?

Marek Czachor*

*Katedra Fizyki Teoretycznej i Metod Matematycznych*
*Politechnika Gdańska, ul. Narutowicza 11/12, 80-952 Gdańsk, Poland*

A classical system violating the Bell inequality is discussed. The system is local, deterministic, observers have free will, and detectors are ideal so that no data are lost. The trick is based on two elements. First, a state of one observer is locally influenced by a "particle". Second, random variables used in the experiment are complementary. A relation of this effect to nonlocality is discussed.

Consider a strip of paper divided into four segments and covered on one side with the following sequence of symbols

$$A+ \quad B'+ \quad A'- \quad B-$$

each of the four symbols appearing in a different segment. The other side of the strip contains the same symbols but with all the signs reversed, i.e.

$$A- \quad B'- \quad A'+ \quad B+$$

and the positions of all the letters with respect to the segments are the same. This means that if we look at a given segment from above and see, say, $A+$ we will see $A-$ if we look at the same segment from below, and so on.

Now let us form a Möbius strip from this piece of paper. The segments $A$ and $A'$ are located antipodally. The same concerns $B$ and $B'$. Such a Möbius strip will play a role of a "particle" in the experiment we will discuss.

Consider now two observers, Alice and Bob, and a "source" which produces pairs of identical Möbius strips. The strips are identically "served" to Alice and Bob. To make this concrete let us assume that the segments which are the closest or the farthest to the person that looks at them are those containing either $A$ or $A'$, so that the $B$ or $B'$ segments are either to the left or to the right with respect to an observer. So, for example, if Alice looks at the strip that has been served and sees $A'-$ in front of her, then she has two possibilities. She either sees $B'+$ at her left and $B-$ at her right or, if the strip is served upside-down, $B-$ at her left and $B'+$ at her right. This is shown in Fig. 1.

Now consider the measurements they make. The orientation shown at Fig. 1 means that Bob finds $B = -1$ and $B' = +1$. He has free will so for a given strip he chooses either $B$ or $B'$. The situation of Alice is assumed to be slightly different. She also has a freedom of measuring either $A$ or $A'$ whatever configuration of the strip is served. For example, when she obtains the configuration from Fig. 1 we say that she is *suggested* to measure $A'$. If she decides to measure $A'$ she just takes the number she can see. If however she decides to measure $A$, she moves clockwise around the strip until she gets to the $A$ segment and then takes the sign she can see. Similarly, she may be suggested to measure $A$ — this depends on the way the strip has been served. We assume that when she rejects a suggestion she moves clockwise on the same side of the strip which she originally saw. In the case from the figure this is the side which contains $A'-$. So for the particular configuration shown at Fig. 1 she gets $A = +1$. A side of the Möbius strip is locally (that is within three subsequent segments) well defined and this is sufficient for the uniqueness of the experiment. Notice, however, that had the strip been served upside-down she would have obtained $A = -1$ (still moving clockwise).

She *accepts* the suggestion with probability $p$ or *rejects* it with probability $1 - p$, but still has free will. She is susceptible to suggestion if $p > 1/2$, totally obedient if $p = 1$, and so on. The probability $p$ is either a feature of Alice or this of an experimental arrangement. We assume that the pairs of strips may be served in all the possible configurations and that all configurations occur with equal probabilities.

The results obtained by Bob are independent of whether the strip is served "normally" or upside-down. The same concerns the result of a suggested Alice experiment. However, whenever Alice rejects the suggestion the results she will obtain are statistically independent of the results obtained by Bob. An acceptance of a suggestion leads to perfect correlations with Bob's results.

Consider now a Bell-type experiment [1]. Both Alice and Bob are allowed to make only one measurement on a given strip. This is similar to an actual situation where for a given pair of photons one measures either $AB$, or $AB'$, or $A'B$, or $A'B'$. The averages for this experiment are

$$\langle A \rangle = \langle A' \rangle = \langle B \rangle = \langle B' \rangle = 0,$$

and

$$\langle AB \rangle = \langle A'B \rangle = \langle AB' \rangle = -\langle A'B' \rangle = 1,$$

whenever Alice accepts a suggestion, and

$$\langle AB \rangle = \langle A'B \rangle = \langle AB' \rangle = \langle A'B' \rangle = 0,$$

whenever she rejects it. The Bell average for an entire experiment is therefore

$$\langle AB \rangle + \langle A'B \rangle + \langle AB' \rangle - \langle A'B' \rangle = 4p.$$

An obedient Alice violates the Bell inequality.

Now where is the secret? The situation is obviously local (Alice and Bob do not communicate), deterministic (once the strips are served the results of all the possible measurements are determined), "detectors" are perfect (no strip is lost), and the observers have free will. Two tricks are used here. First, a state of Alice is *locally* influenced by the "particle". Second, the observables $A$ and $A'$, and $B$ and $B'$, are *complementary*. To understand this consider the situation from the figure. If Alice decides to measure first $A'$ and then $A$ she obtains $A' = -1$ and $A = +1$. But if she measures first $A$ and then $A'$ she finds $A = +1$ and $A' = +1$ and the two measurements do not commute. This is why we have chosen the Möbius strip and not a cylinder. The results of $A$ and $A'$ cannot be simultaneously considered because they depend on the order of measurements. Therefore the random variable $A \pm A'$ makes no sense and the Bell inequality cannot be derived. This element is also technically responsible for the fact that Bob's freedom does not have to be limited in any way (his $p = 1/2$).

A few additional comments on the role of $p$ have to be made. Imagine that when Alice is served a plate with the Möbius strip the plate is put either at her right or at her left. To measure the random variable which is suggested she just looks at the strip. To reject the suggestion she has to rotate the strip and look what sign is at the opposite side of the strip. To do so she performs the rotation with her right arm if the strip is served at her right and with her left arm in the opposite case. But Bob's waiter puts the strip always just in front of him so that he has completely no problem with deciding where to look.

In a long run, say after some 100000 dishes served during this dinner, it may turn out that Alice more often accepts a suggestion which is served at her left even though she is not aware of it. The reason is that her left arm is more susceptible to fatigue, a condition typical of right-handed persons. The parameter $p$ is here determined after the experiment is completed. One can even check experimentally whether she is right-handed by measuring the Bell average!

For quantum mechanical purists I can reformulate this story as follows. Imagine that we have a source of pairs of particles. The particles are emitted in two cones whose spherical angles are $2\epsilon$. Particles from one cone are collected by Alice and the other ones by Bob. Alice has analyzers which can collect particles emitted in a cone of angle $\epsilon$, but Bob uses a newer model which collects those emitted in a cone of $2\epsilon$. So Alice has to use two analyzers placed next to each other whereas Bob can use a single one. Now she operates one of them with her right hand and the other with her left hand, and the analysis given above can be repeated. Notice that an element which must be included is that the Alice particles locally *try* to change a state of her analyzers and it is her free will that finally makes that, to the extent given by $p$, (im)possible.

One can easily see how a nonlocal communication between Bob and Alice can enhance the violation. Assume that after Bob decides which observable to measure he communicates this to Alice. Once she knows this, she knows also which path (clockwise or counter-clockwise) to choose to get from the suggested sector to the opposite one in order to obtain the maximal violation. Here complementarity comes again but nonlocally: The result of $A$ depends on the trajectory on the strip i.e. the decision made by Bob.

This paper is a result of several violent conversations on the subject I had with Paweł Horodecki, Michał Horodecki and Marek Żukowski. Also discussions I had a few years ago with Lev Vaidman and Lucien Hardy left me with a feeling that a clear counter-example has to be found. I gratefully acknowledge remarks of all of them. I am idebted also to Magda Zarzycka who helped me to clarify some points. The work is a part of the Polish-Flemish Project No. 007.

---

∗ Electronic address: mczachor@sunrise.pg.gda.pl
[1] J. S. Bell, Physics **1**, 195 (1964).